\documentclass[twocolumn,english,aps,pra,twocolum,superscriptaddress,natbib,amsmath,amssymb,floatfix,superscriptaddress,footinbib]{revtex4-2}
\usepackage[colorlinks=true,citecolor=blue,linkcolor=magenta]{hyperref}
\usepackage{color} 

\newcommand{\blue}[1]{\textcolor{black}{#1}}
\usepackage[addedmarkup=colored, authormarkupposition=left]{changes} 

\usepackage{soul}
\definechangesauthor[name={TJK}, color=red]{TJK}
\definecolor{hugoColor}{RGB}{59,134,255}
\definechangesauthor[name={HL}, color=hugoColor]{HL}

\usepackage[english]{babel}
\usepackage{booktabs}
\usepackage{amsmath,amsfonts,amssymb}
\usepackage[T1]{fontenc}
\usepackage{xurl}

\usepackage{amsmath}
\usepackage{siunitx}
\usepackage[version=4]{mhchem}

\DeclareSIUnit \baud {Bd}

\usepackage{amsfonts}
\usepackage{amssymb}
\usepackage{epstopdf}
\usepackage{graphicx}
\usepackage{float}
\graphicspath{{./Figures/}}
\usepackage[utf8]{inputenc}
\UseRawInputEncoding

\newcommand{\threefive}[0]{\uppercase\expandafter{\romannumeral3}-\uppercase\expandafter{\romannumeral5}~}

\begin{document}
    \title{Lithium tantalate integrated extended grating-based lasers with Hertz linewidth}
    
    \author{Hugo~Larocque}
    \thanks{These authors contributed equally.}
    \affiliation{Institute of Physics, Swiss Federal Institute of Technology Lausanne (EPFL), CH-1015 Lausanne, Switzerland}
    \affiliation{Institute of Electrical and Micro engineering, Swiss Federal Institute of Technology, Lausanne (EPFL), CH-1015 Lausanne, Switzerland}
    
    \author{Zhuoya~Yuan}
    \thanks{These authors contributed equally.}
    \affiliation{Institute of Physics, Swiss Federal Institute of Technology Lausanne (EPFL), CH-1015 Lausanne, Switzerland}
    \affiliation{Institute of Electrical and Micro engineering, Swiss Federal Institute of Technology, Lausanne (EPFL), CH-1015 Lausanne, Switzerland}

    \author{Zihan~Li}
    \affiliation{Institute of Physics, Swiss Federal Institute of Technology Lausanne (EPFL), CH-1015 Lausanne, Switzerland}
    \affiliation{Institute of Electrical and Micro engineering, Swiss Federal Institute of Technology, Lausanne (EPFL), CH-1015 Lausanne, Switzerland}

    \author{Giovanni~Scarioni}
    \affiliation{Institute of Physics, Swiss Federal Institute of Technology Lausanne (EPFL), CH-1015 Lausanne, Switzerland}
    \affiliation{Institute of Electrical and Micro engineering, Swiss Federal Institute of Technology, Lausanne (EPFL), CH-1015 Lausanne, Switzerland}

    \author{Annika~Schoels}
    \affiliation{Institute of Physics, Swiss Federal Institute of Technology Lausanne (EPFL), CH-1015 Lausanne, Switzerland}
    \affiliation{Institute of Electrical and Micro engineering, Swiss Federal Institute of Technology, Lausanne (EPFL), CH-1015 Lausanne, Switzerland}

    \author{Jiale~Sun}
    \affiliation{Institute of Physics, Swiss Federal Institute of Technology Lausanne (EPFL), CH-1015 Lausanne, Switzerland}
    \affiliation{Institute of Electrical and Micro engineering, Swiss Federal Institute of Technology, Lausanne (EPFL), CH-1015 Lausanne, Switzerland}

    \author{Anat~Siddharth}
    \affiliation{Deeplight SA, St-Sulpice CH-1025, Switzerland}

    \author{Xin~Ou}
    \affiliation{State Key Laboratory of Materials for Integrated Circuits, Shanghai Institute of Microsystem and Information Technology, Chinese Academy of Sciences, Shanghai, China}

    \author{Simone~Bianconi}
    \email[]{simone.bianconi@epfl.ch}
    \affiliation{Institute of Physics, Swiss Federal Institute of Technology Lausanne (EPFL), CH-1015 Lausanne, Switzerland}
    \affiliation{Institute of Electrical and Micro engineering, Swiss Federal Institute of Technology, Lausanne (EPFL), CH-1015 Lausanne, Switzerland}
    
    \author{Tobias~J.~Kippenberg}
    \email[]{tobias.kippenberg@epfl.ch}
    \affiliation{Institute of Physics, Swiss Federal Institute of Technology Lausanne (EPFL), CH-1015 Lausanne, Switzerland}
    \affiliation{Institute of Electrical and Micro engineering, Swiss Federal Institute of Technology, Lausanne (EPFL), CH-1015 Lausanne, Switzerland}

\maketitle
\noindent 

    
     \noindent \textbf{Lasers that combine narrow linewidth with fast, wide-range frequency tuning underpin coherent light detection and ranging (LiDAR), distributed fiber sensing, microwave photonics and spectroscopy. Hybrid integrated lasers have recently demonstrated lower phase noise than the best legacy fiber lasers, while being compact and scalable to manufacture. Recent advances in thin-film lithium niobate have enabled hybrid external cavity-based lasers with gigahertz-tuning range at hundred-megahertz rates. Nonetheless, the lithium niobate platform presents some drawbacks, such as photorefractive effects and bias drift. Moreover, the intrinsic linewidth of these Pockels lasers (kHz) remains well above those of other integrated lasers: hence fast tuning comes at the cost of higher noise.
    Here, we show that e-DBR lasers in thin-film lithium tantalate can overcome this trade-off, reaching intrinsic linewidths as low as 24 Hz (1.9 kHz integrated linewidth), seven times lower than the best lithium niobate e-DBR lasers. The e-DBR lasers were fabricated with deep-ultraviolet lithography on 4-inch lithium-tantalate-on-insulator wafers and characterized across the full wafer, exhibiting a statistically significant noise reduction compared to its lithium niobate-based counterparts. Crucially, Pockels tuning is fully retained: 2 GHz chirps at up to 1 MHz for 5 V peak-to-peak drive, with $\pm\text{30}$ MHz linearity error, a 190 MHz electro-optic bandwidth and, owing to the low electro-optic drift of lithium tantalate, uniform tuning down to tens-of-hertz drive frequencies. We showcase a butterfly-packaged laser operating in a free-running state within a 232 MHz band over 19 hours and performing coherent LiDAR of a live scene. Together with lithium tantalate's mature supply chain from acoustic-filter manufacturing, wider bandgap and copper-damascene compatibility for electronic co-integration, these results establish a scalable platform for ultrafast, widely tunable and narrow-linewidth integrated lasers.}
    \section{Introduction} 
 

   Narrow-linewidth, frequency-agile lasers are key to several established and emerging systems, such as those for light detection and ranging (LiDAR)~\cite{h.bostickCarbonDioxideLaser1967}, distributed-fiber sensing~\cite{luDistributedOpticalFiber2019}, and microwave photonics~\cite{marpaungIntegratedMicrowavePhotonics2019b}. Integrated photonic lasers can provide a unique combination of unprecedented narrow-linewidth and high-speed tunability over a compact footprint~\cite{lihachev_Lownoise_2022}, becoming a key enabler for many of these emerging technologies~\cite{wangHighPerformanceIntegratedLaser2024}. Among available integrated laser architectures, extended distributed Bragg reflector (e-DBR) lasers distinguish themselves by their output powers in the tens of mW and wide tuning ranges~\cite{tran_Tutorial_2019,huang_Highpower_2019,xiang_Ultranarrow_2019,siddharth_Piezoelectrically_2024}. Implementing such devices in an integrated platform based on materials featuring a strong Pockels effect~\cite{li_High_2023,wang_Lithium_2024} allows them to benefit from highly efficient GHz-rate electro-optic modulation without undergoing additional optical losses. For instance, e-DBR lasers manufactured in thin-film lithium niobate yield tuning ranges exceeding \SI{10}{\giga\hertz} and tuning rates near \SI{200}{\mega\hertz}, thereby making them instrumental in achieving cm-scale resolution in LiDAR and tuning ranges sufficient for practical spectroscopy~\cite{siddharth_Ultrafast_2025,xue_Pockels_2025}. However, producing such Pockels e-DBR lasers with sufficiently low noise remains an open challenge.

   Implementing these lasers in alternative Pockels materials could provide a path towards further reducing their noise. For instance, lithium tantalate integrated photonic devices often feature levels of photostability beneficial for some practical tasks like optical modulation~\cite{lin_Copper_2026}, considerable promise as a material suitable for volume manufacturing~\cite{wang_Lithium_2024}, while also showing compatibility with visible photonics due to its larger bandgap~\cite{meynTunableUltravioletRadiation1997, katz_Vaportransport_2004a, hum_quasiphasematched_2007}. Early demonstrations of hybrid integrated lithium tantalate e-DBR lasers already suggest that the material's stability can lead to narrow linewidths~\cite{xue_narrow_2025}. This feature, along with ultrafast modulation and volume manufacturing compatibility observed in other photonic integrated devices~\cite{wang_Lithium_2024}, potentially position lithium tantalate as a leading candidate for readily deployable frequency-agile narrow linewidth lasers. Here, we demonstrate lithium tantalate's full lasing potential with integrated devices combining narrow linewidth with fast and efficient Pockels tuning. By means of wafer-scale manufacturing and exhaustive testing of the fabricated devices, we demonstrate integrated lasers with an intrinsic linewidth reaching values as low as \SI{24}{\hertz}. This value is considerably lower than what lithium niobate devices can currently achieve without compromising on tuning efficiency nor bandwidth.

    \begin{figure*}[htbp!]
		\centering
            \includegraphics[width=1\linewidth]{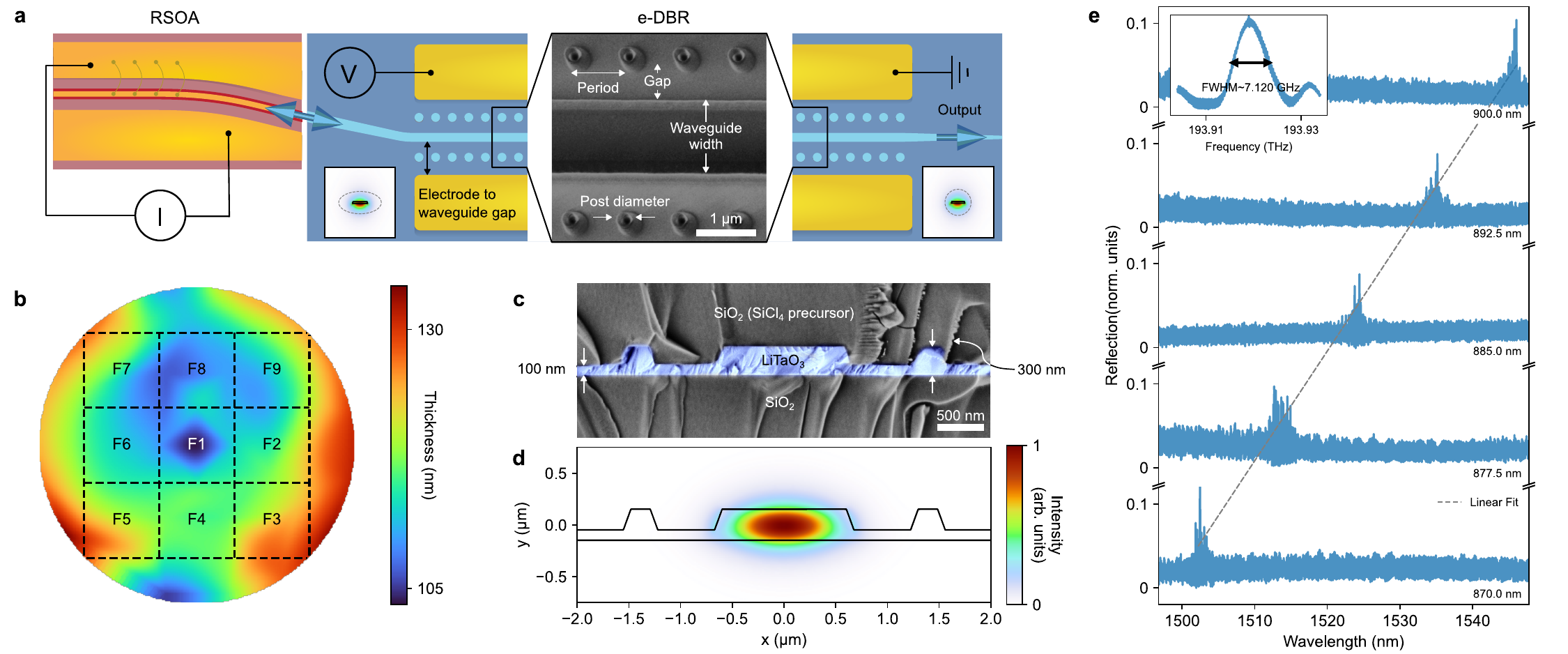}
		\caption{\textbf{Lithium tantalate distributed Bragg reflector.} \textbf{a,} Schematics of the lithium tantalate e-DBR hybrid laser. The device consists of a reflective semiconductor optical amplifier butt coupled to the lithium tantalate PIC and a lensed fiber collecting the laser's emission. Insets: mode profiles of the input and output waveguides at their corresponding facets. The dashed lines indicate the half-maximum intensity contours of the modes coupling to the waveguides, specifically the near-field modes of the hybrid laser's RSOA and lensed fiber; Scanning electron micrograph of an uncladded lithium tantalate Bragg reflector relying on a post grating structure. \textbf{b,} Thickness map of the processed 4-inch wafer (Wafer ID: D010191) following the main lithium tantalate patterning step of the grating's fabrication. The map specifically consists of interpolated data calculated from 39 optical reflectometer measurements acquired in flat areas across the wafer. \textbf{c,} False-colored cross-section scanning electron micrograph of the grating shown in \textbf{a}. \textbf{d}, Simulated profile of the optical mode reflected by the grating. \textbf{e}, Reflection spectra of post gratings with periods ranging from 870~nm to 900~nm. The device features a \SI{1.3}{\micro\meter} nominal waveguide width, a \SI{375}{\nano\meter} post diameter, and a \SI{500}{\nano\meter} post-to-waveguide gap. Inset: Main lobe of the \SI{900}{\nano\meter} period DBR reflection spectrum. Figure legend: RSOA: reflective semiconductor optical amplifier, e-DBR: extended distributed Bragg reflector.}
		\label{fig:fig1}
	\end{figure*}


    \section{Results} 

    Figure~\ref{fig:fig1}{a} provides the schematics for the lithium tantalate PIC within the hybrid laser. Its primary feature consists of a post grating acting as the laser's DBR. Electrodes allow for tuning its reflection spectrum by means of the Pockels effect. Waveguides book-ending the grating at the PIC's facets have a tapering profile enabling mode matching with the hybrid architecture's other photonic components. The fabrication of the integrated laser starts with a 4-inch X-Cut lithium tantalate on insulator wafer. The wafer holds a 300~nm thick lithium tantalate thin film prepared with a hydrogen-based ion slicing technique~\cite{yan_WaferScale_2019,wang_Lithium_2024} and bonded on a silicon substrate with \SI{4.7}{\micro\meter} of buried oxide. Following Deep-UV lithography, an argon ion physical etch patterns the membrane down to roughly \SI{100}{\nano\meter}. Both the uniformity of the starting film and of the fabrication process introduce variations around these nominal thicknesses. To account for these variations, Fig.~\ref{fig:fig1}{b} provides a wafer map of the lithium tantalate film thickness following its patterning across the nine DUV stepper fields used in the fabrication. Each field includes a copy of the PIC used in this work. The fabrication concludes with the deposition of a silicon dioxide cladding using inductively coupled plasma chemical vapor deposition and metallization~\cite{li_High_2023}. 
    
    As depicted in the scanning electron micrograph (SEM) inset of Fig.~\ref{fig:fig1}{a}, the fabricated lithium tantalate grating has a nominal \SI{1.3}{\micro\meter}-wide waveguide flanked by posts with a \SI{375}{\nano\meter} diameter. These structural parameters define all the gratings explored in this work. The grating shown in the micrograph specifically features a post-to-waveguide gap of \SI{600}{\nano\meter}, as more readily observed in the cross-sectional SEM of the grating provided in Fig.~\ref{fig:fig1}{c}. For reference, Fig.~\ref{fig:fig1}{d} shows the corresponding guided mode profile in the grating, highlighting that its posts minimally perturb the optical mode propagating through the grating. This design feature targets low coupling DBR structures, ensuring narrow-linewidth reflection spectra, suitable for lasing, at the cost of lower reflectivity~\cite{spencer_Low_2015}. Nonetheless, these grating resonances clearly appear in the gratings' reflection spectra displayed in Fig.~\ref{fig:fig1}{e}. The five considered grating periods feature reflection peaks spanning roughly \SI{50}{\nano\meter}, thus highlighting this platform's ability to produce devices suitable for lasing at various wavelengths. As indicated by the dashed line, the resonances follow the expected linear relation $\lambda_0= 2n_\mathrm{eff}\Lambda$, where $n_\mathrm{eff}$ is the guided mode's effective index and $\Lambda$ is the grating period.

    \begin{figure*}[htbp!]
		\centering
            \includegraphics[width=1\linewidth]{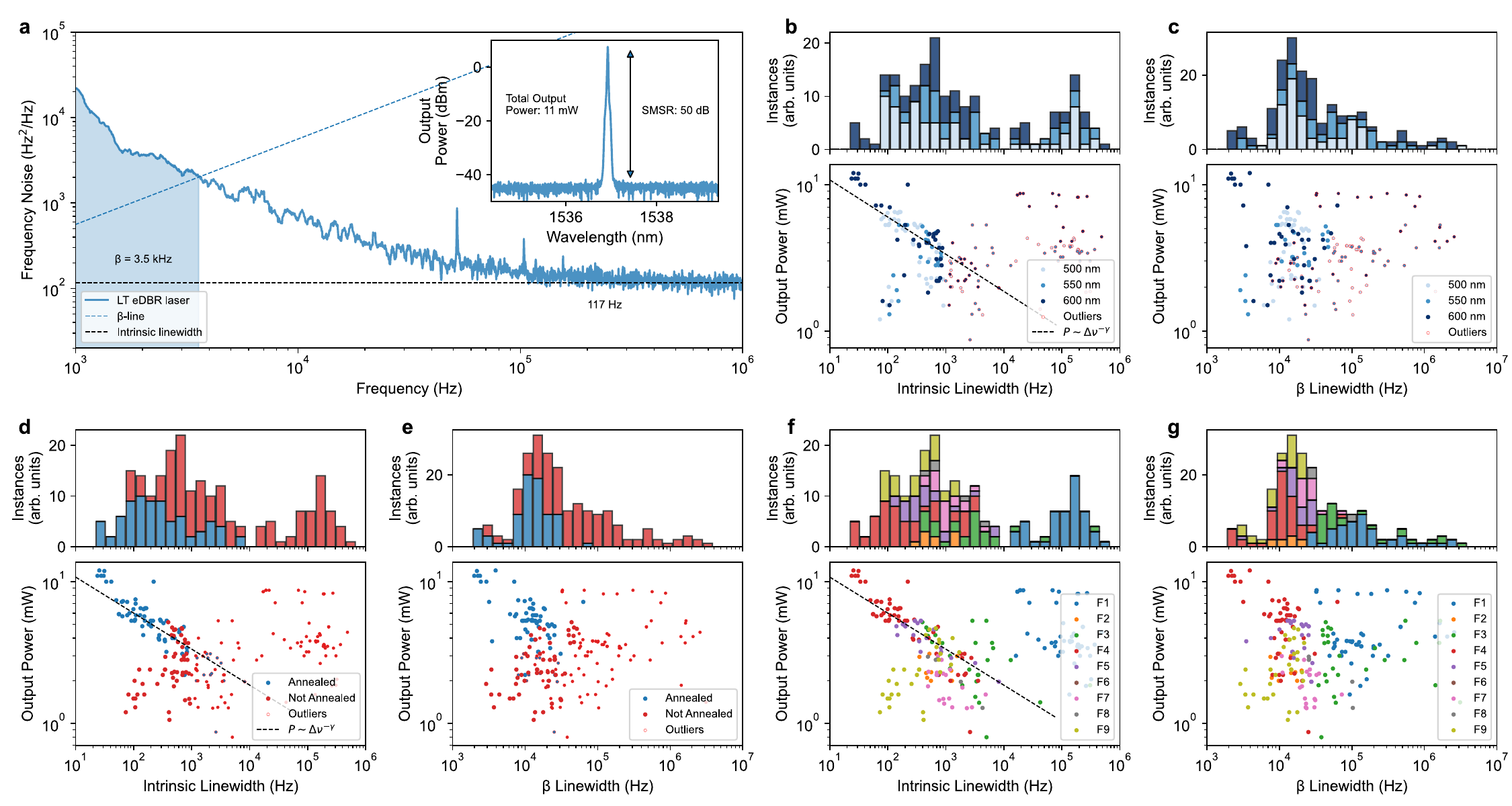}
		\caption{\textbf{Spectral features of lithium tantalate extended-DBR integrated lasers.} \textbf{a,} Representative frequency noise spectrum of the lithium tantalate extended-DBR laser consisting of a grating with a \SI{1.3}{\micro\meter} nominal waveguide width, a \SI{375}{\nano\meter} post diameter, a \SI{600}{\nano\meter} post-to-waveguide gap, and a \SI{892.5}{\nano\meter} period (Field 5, Chip 7, Waveguide 4.03). Inset: emission spectrum of a lithium tantalate extended-DBR laser. The spectrum features a resolution of \SI{0.02}{\nano\meter} \textbf{b,} Intrinsic and \textbf{c,} integrated beta linewidth statistical analyses for extended-DBR lasers over a set of devices operating under various conditions. Beta linewidth measurements shown here have a \SI{1}{\milli\second} integration time. The device post-to-waveguide spacing determines the color coding of the plotted data. Top row: histograms plotting the occurrence of lasing states defined by specific linewidths. Bottom row: scatter plots showing the instances from the histograms and the corresponding fiber-coupled output power or the intrinsic linewidth of the integrated laser. The dashed line represents a linear fit of the output power against the inverse of the intrinsic linewidth. \textbf{d,} Intrinsic and \textbf{e,} beta linewidth data shown in \textbf{b,c} colored based on whether the lithium tantalate PIC experienced an additional annealing step in its fabrication. \textbf{f,} Intrinsic and \textbf{g,} beta linewidth data shown in \textbf{b,c} colored based on the field number of the device as defined in Fig.~\ref{fig:fig1}d. Figure legend: SMSR: side-mode suppression ratio.}
		\label{fig:fig2}
	\end{figure*}

    Building on the fabrication method and spectral features of the lithium tantalate DBR illustrated in Fig.~\ref{fig:fig1}, we proceed by demonstrating its suitability for lasing. As illustrated in Fig.~\ref{fig:fig1}a, we rely on a hybrid configuration involving a reflective semiconductor optical amplifier (RSOA, Thorlabs SAF1126c) butt-coupled to the DBR with a lensed fiber collecting the resulting e-DBR laser's output~\cite{xiang_Ultranarrow_2019,siddharth_Piezoelectrically_2024, siddharth_Ultrafast_2025,xue_Pockels_2025}. 
    To perform linewidth measurements on our hybrid lasers' emission, we employ a delayed self-heterodyne method relying on a Wiener filter model to recover the lasers' frequency noise spectrum~\cite{kantner2023accurate,riebesehl_Interference_2025}. Figure~\ref{fig:fig2}{a} exemplifies the results of this procedure with a representative noise spectrum. This particular device defined by a \SI{500}{\nano\metre} post-to-waveguide gap features intrinsic linewidth noise floor and integrated ($\beta$) linewidth of \SI{117}{\hertz} and \SI{3.5}{\kilo\hertz}, respectively. The inset in Fig.~\ref{fig:fig2}{a} further provides the laser's corresponding emission spectrum featuring an output power of \SI{5.7}{\milli\watt} and a side mode suppression ratio of 50~dB. 
    To better quantify how various device and experimental parameters affect the underlying laser's linewidth, we provide an exhaustive account of linewidth measurements in Figs.~\ref{fig:fig2}{b,c}. These measurements consider factors ranging from post-to-waveguide spacing in the grating's design to electrical current values for pumping and altering the gain properties of the RSOA. 
    They also account for multiple copies of the same device design manufactured on the lithium tantalate wafer shown in Fig.~\ref{fig:fig1}{b}, thereby incorporating potential effects related to process drift and fabrication variations affecting device performance. The intrinsic and $\beta$-linewidth histograms of Fig.~\ref{fig:fig2}{b} and Fig.~\ref{fig:fig2}{c} suggest the underlying distributions of both variables peak near \SI{530}{\hertz} and \SI{12}{\kilo\hertz}, respectively, with one data point showing that our platform can reach values as low as \SI{24}{\hertz} and \SI{1.9}{\kilo\hertz}. Our considered post-to-waveguide gap values do not significantly affect the underlying linewidth distribution of the resulting lasers. To illustrate the potential influence of other experimental parameters on laser linewidth, we also plot the points considered in the intrinsic linewidth histogram against their corresponding laser's output power. Excluding data points attributed to samples on stepper fields that yielded noisy lasers, we observe that the lasers' output power scales inversely with their measured intrinsic linewidths~\cite{schawlow_Infrared_1958}. 

    PICs from some of the DUV stepper fields shown in Fig.~\ref{fig:fig1}{b} experienced an additional 2-hour long $500^\circ$C annealing processing step after their fabrication. To correlate measured linewidths with this factor, Figs.~\ref{fig:fig2}{d,e} provides a version of Figs.~\ref{fig:fig2}{b,c} which colors data points based on whether they correspond to annealed samples. Likewise, Figs.~\ref{fig:fig2}{f,g} color the data based on the sample's stepper field. 

\begin{figure*}[htbp!]
		\centering
            \includegraphics[width=1\linewidth]{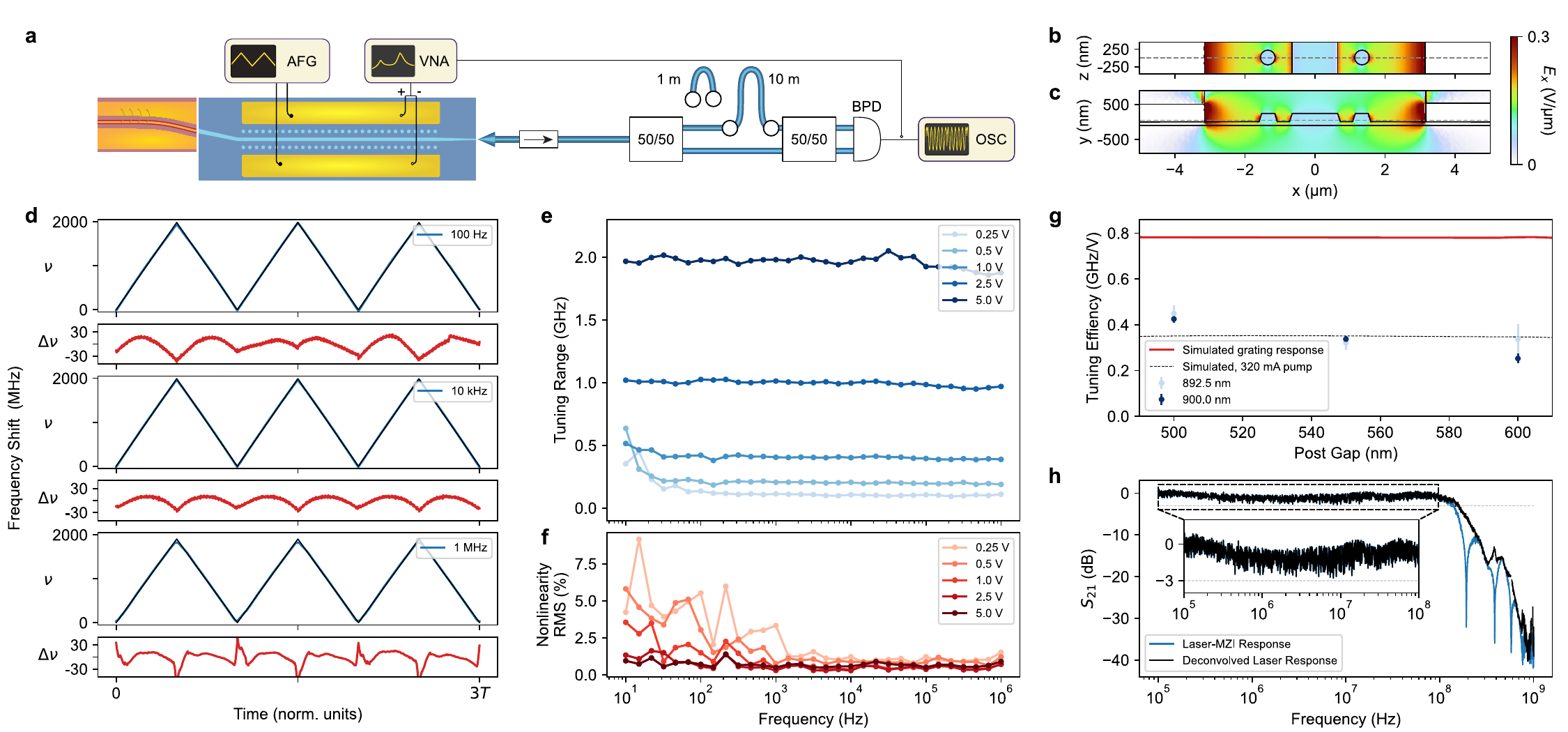}
		\caption{\textbf{Tuning features of lithium tantalate extended-DBR integrated lasers.} \textbf{a,} Experimental apparatus for measuring wavelength tuning achieved in lithium tantalate e-DBR lasers. An arbitrary function generator tunes the laser which sends its emission into a Mach-Zehnder interferometer with a 10~m long delay fiber. An oscilloscope then reads out the collected signal from a balanced photodetector for data acquisition. \textbf{b,} Top and \textbf{c,} cross-section views of the electric field applied across a unit cell of the e-DBR laser to tune its emission. \textbf{d,} Measured optical frequency shift in the integrated laser spectrum while applying a 5~Vpp triangle wave to the lithium tantalate grating at \SI{100}{\hertz}, \SI{10}{\kilo\hertz}, and \SI{1}{\mega\hertz} drive frequencies. The laser's grating features a \SI{1.3}{\micro\meter} nominal waveguide width, a \SI{375}{\nano\meter} post diameter, a \SI{500}{\nano\meter} post-to-waveguide gap, and a \SI{900}{\nano\meter} period. The plotted error signals, $\Delta \nu$, correspond to the difference between the measured data and the perfect triangle waves plotted as black lines. \textbf{e,} Maximum tuning range of the integrated lasers while applying triangle drive signals with amplitudes ranging from 0.25 to 5~Vpp and frequencies from \SI{10}{\hertz} to \SI{1}{\mega\hertz}. The laser's grating features the same parameters as the ones presented in \textbf{a}. \textbf{f,} Root-mean square of the deviation between the measured frequency shifts and ideal triangle waves for the data shown in \textbf{b}. \textbf{g,} Average tuning efficiency of various lasing states for lasers featuring post-to-waveguide gaps of \SI{500}{\nano\meter}, \SI{550}{\nano\meter}, and \SI{600}{\nano\meter}. The rest of the parameters include a \SI{1.3}{\micro\meter} nominal waveguide width, \SI{375}{\nano\meter} post diameter, along with \SI{892.5}{\nano\meter} and \SI{900}{\nano\meter} periods. The dashed black line corresponds to the tuning efficiency expected from finite element simulations. \textbf{h,} Electrical-to-optical $\mathrm{S}_{21}$ parameter obtained after replacing the function generator and oscilloscope in \textbf{a} by a vector network analyzer and shortening the 10~m delay down to 1~m. Figure legend: AFG: arbitrary function generator, BPD: balanced photodetector, OSC: oscilloscope, VNA: vector network analyzer.}
		\label{fig:fig3}
	\end{figure*}
    
    Implementing an e-DBR laser within lithium tantalate also introduces the prospect of ultrafast emission wavelength tuning enabled by this material's Pockels effect. To verify this capability, we endow the e-DBR PICs with electrodes capable of applying an electric field across the grating and employ the apparatus depicted in Fig.~\ref{fig:fig3}a. As depicted in Fig.~\ref{fig:fig1}a, gaps of \SI{2.5}{\micro\meter} separate the electrodes from the edge of the grating's nominal waveguide. The orientation of this structure ensures that the electric field points along the crystal's extraordinary axis, thereby setting lithium tantalate's largest Pockels tensor component, $r_{33}=30.5$~pm/V, as the main tuning contributor. Figures~\ref{fig:fig3}{b},{c} provide top and cross-section views of the tuning electric field that result from applying \SI{1}{\volt} across the electrodes. The plotted component corresponds to the one aligned with the extraordinary axis of the lithium tantalate film.
    
    To rapidly detect induced variations in the laser's emission wavelength, we rely on the interferometric apparatus shown in Fig.~\ref{fig:fig3}{a}. 
    Figure~\ref{fig:fig3}{d} shows the result of such tuning on a laser's optical frequency, $\nu$, while applying a 5~Vpp triangle wave to the e-DBR's electrodes. This specific laser has a grating design defined by a \SI{500}{\nano\meter} post-to-waveguide gap and a \SI{900}{\nano\meter} period. For drive frequencies of \SI{100}{\hertz}, \SI{10}{\kilo\hertz}, and \SI{1}{\mega\hertz}, the laser displays frequency chirps spanning near \SI{2}{\giga\hertz} with deviations, $\Delta \nu = \nu-\nu_\mathrm{linear}$, from an ideal triangle wave $\nu_\mathrm{linear}$ constrained to within roughly $\pm30$~MHz. We repeat this measurement over a larger set of waveforms with frequencies between \SI{10}{\hertz} and \SI{1}{\mega\hertz} and amplitudes ranging from \SI{0.25}{\volt} to \SI{5}{\volt}. Figure~\ref{fig:fig3}{e} plots the tuning range, $\max({\nu})$, resulting from applying these waveforms to the e-DBR. This range linearly increases with the amplitude of the applied signal, as expected from the linear nature of the Pockels effect. The tuning features good uniformity across most of the considered frequency range. Below frequencies of \SI{100}{\hertz}, the maximum tuning begins to stray from its nominal high frequency value, which is an indication of the low frequency drift commonly observed in ferroelectric materials such as lithium niobate~\cite{holzgrafeRelaxationElectroopticResponse2024} and tantalate~\cite{lin_Copper_2026}. More pronounced deviations from perfectly linear frequency tuning also reflect this trait. Figure~\ref{fig:fig3}{f} emphasizes these results by plotting the root mean square of the error signal, $\Delta \nu_\mathrm{rms}$, attributed to this deviation. Overall, these results show that the deviations become more pronounced for low frequency drive signals and those applying weaker electric fields across the device. To better illustrate these effects, Supplementary Note 1 provides sample traces of such frequency ramps under various drive frequencies and amplitudes.

    \begin{figure*}[htbp!]
		\centering
            \includegraphics[width=1\linewidth]{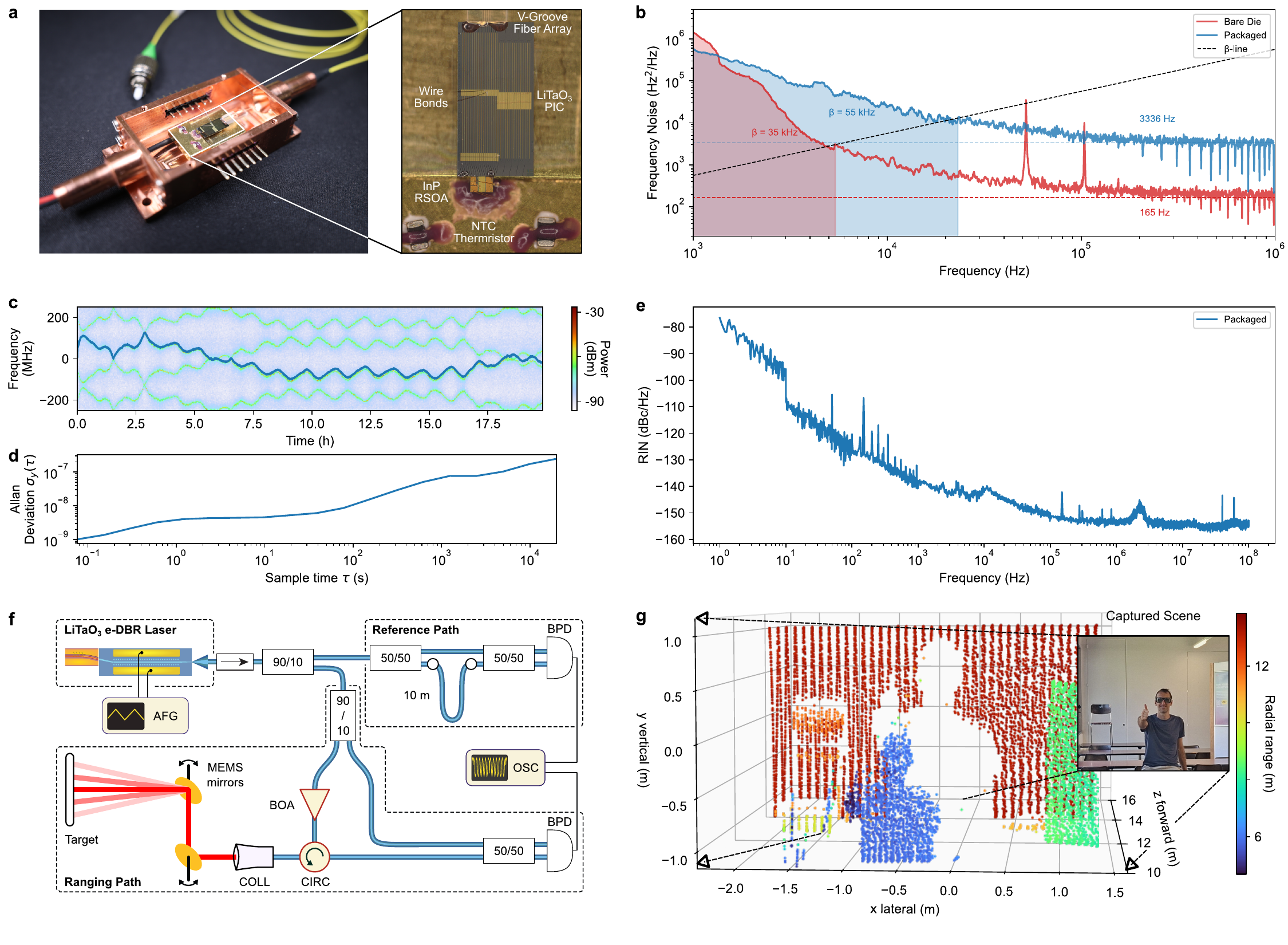}
		\caption{\textbf{Packaging and deployment of a lithium tantalate extended-DBR laser.} \textbf{a,} Photograph of the packaged lithium tantalate extended-DBR laser. Inset: micrograph of the packaged laser highlighting the various dies and components used in the assembly. \textbf{b,} Frequency noise of extended-DBR lasers with and without packaging relying on the same lithium tantalate DBR (Field 4, \SI{500}{\nano\meter} post-gap spacing, \SI{900}{\nano\meter} period). \textbf{c,} Time-frequency spectrogram of the packaged laser obtained by beating its emission to a 250~MHz repetition rate frequency comb (Menlo Systems SmartComb-250) overlaid with the extracted measured laser frequency drift. \textbf{d,} The corresponding Allan deviation of the laser's emission based on the spectrogram from \textbf{c}. \textbf{e,} Relative intensity noise (RIN) of the packaged laser. \textbf{f,} Apparatus for testing the hybrid laser's suitability for light detection and ranging. \textbf{g,} Captured point cloud attributed to the measured depth map of a live scene. Figure legend: AFG: arbitrary function generator, BPD: balanced photodetector, BOA: booster semiconductor optical amplifier, CIRC: optical circulator, COLL: collimator, MEMS: micro electro-mechanical systems.}
		\label{fig:fig4}
	\end{figure*}
    
    We repeat such experiments on e-DBRs defined by different grating parameters. Specifically, we drive a set of six lasers at a \SI{10}{\kilo\hertz} drive frequency with 892.5-900~nm period gratings and 500-600~nm post-to-waveguide gaps while considering various currents applied to the RSOA. Figure~\ref{fig:fig3}{g} plots the average tuning range of devices sharing the same post-to-waveguide gap, which implies that increasing the grating's gap marginally decreases tuning efficiency. These results line up with simulated data obtained from a traveling wave model considering propagation within the laser's hybrid cavity formed by the III-V gain chip and the lithium tantalate grating chip~\cite{t.smyIntegrationTravelingWave2020, j.h.rasmussenImplementationTravelingWave2023}. These results additionally consider the expected shift in the grating's reflection spectrum, which we obtain from perturbative calculations relying on electrostatic and eigenmode finite element simulations~\cite{larocque_Photonic_2024,cai_Heterogeneously_2026}. For reference, Fig.~\ref{fig:fig3}{g} also plots this grating response. Supplementary Note~2 provides additional details regarding the methodology behind these simulations.

    To quantify the tuning bandwidth of the lithium tantalate e-DBR, we measure its electrical-to-optical $\mathrm{S}_{21}$ parameter while driving the electrodes with a -17~dBm small signal and monitoring its output with a 1~GHz bandwidth detector. As shown in Fig.~\ref{fig:fig3}h, this measurement suggests a 3~dB bandwidth of \SI{190}{\mega\hertz}.

    We proceed by gauging the laser's durability and its capabilities for prospective applications. To ensure long-term operation, we assemble our hybrid laser in a butterfly package. Figure~\ref{fig:fig4}a shows the resulting device packaged using the methodology and schematics outlined in Supplementary~Note~3. Figure~\ref{fig:fig4}b provides the frequency noise of two lasers relying on the same lithium tantalate PIC. The first operates in a configuration where mounting on multi-axis flexure stages ensures optical coupling between the laser's constituting components, whereas the second operates in the packaged configuration shown in Fig.~\ref{fig:fig4}a. Experimental conditions for both measurements include pumping currents of 170~mA and 140~mA with monitored output powers of 1.86~mW and 4.6~mW for the bare-die and packaged configurations, respectively. We then beat its output with a locked \SI{250}{\mega\hertz} repetition rate frequency comb to measure deviations in its emission frequency while operating in a free-running state. Figure~\ref{fig:fig4}c provides the resulting spectrogram overlaid with the frequency drift of one of its beat notes. We observe drifting within a range of \SI{232}{\mega\hertz} over a duration of 19 hours. Figure~\ref{fig:fig4}d correspondingly shows the Allan deviation obtained from the spectrogram, thus indicating the magnitude of various noise contributions occurring over a range of time scales. With the frequency stability properties of the laser established, we proceed by measuring the relative intensity noise of the packaged laser. Figure~\ref{fig:fig4}e shows the results of this measurement, with Supplementary~Note~4 providing additional methodological details and comparisons with a similarly packaged lithium niobate e-DBR laser.

    Finally, we benchmark the tuning capabilities of our lithium tantalate e-DBR laser in a Pockels laser-based LiDAR demonstration~\cite{siddharth_Ultrafast_2025, xue_Pockels_2025}. Figure~\ref{fig:fig4}f illustrates the schematics of the ranging apparatus. Figure~\ref{fig:fig4}g shows the acquired point cloud providing depth information on a monitored scene. We provide additional details on the extraction of this map from the captured photodiode signals in Supplementary~Note~5.

    \section{Discussion} 
  
    Close material similarities between lithium tantalate and lithium niobate~\cite{wang_Lithium_2024} warrant a comparison between e-DBR lasers implemented in these two material platforms. Supplementary~Note~6 provides a summary of their key performance metrics. Prior reports~\cite{siddharth_Ultrafast_2025, xue_Pockels_2025} indicate output powers readily achievable in our platform, whereas reported intrinsic linewidths can lie anywhere between \SI{167}{\hertz} and \SI{2.5}{\kilo\hertz}. Our best lithium tantalate-based device lies considerably below this range. In comparison with previously demonstrated cladding-defined lithium tantalate grating-based lasers~\cite{xue_narrow_2025}, we confirm that our directly patterned lithium tantalate-based e-DBR lasers further exhibit high-speed, efficient Pockels frequency tuning. Recent work reporting good photostability in integrated lithium tantalate modulators~\cite{wang_Lithium_2024, lin_Copper_2026, cai_Heterogeneously_2026} could suggest that this feature is conducive to improved noise in hybrid lasers. However, these improved figures could also result from the statistical nature of our exhaustive wafer-scale device measurements. As observed in Fig.~\ref{fig:fig2}{b}, lithium tantalate microfabrication can provide suboptimal devices with intrinsic linewidths covering the range of those achieved in lithium niobate. Based on the wafer map in Fig.~\ref{fig:fig1}{b} and the data from Fig.~\ref{fig:fig2}{f}, these variations likely arise from altering wafer-level device geometries that can affect factors pertaining to laser linewidth such as the reflection spectrum of the DBR and the coupling efficiency between the RSOA and the lithium tantalate PIC. 
    To illustrate, field F1 features a thickness profile and an emission linewidth distribution that both significantly stray from those of the other stepper fields.
    Such factors also likely justify the varying figures achieved in works relying on the same material platform~\cite{xiang_Ultranarrow_2019,siddharth_Piezoelectrically_2024,xue_Pockels_2025,siddharth_Ultrafast_2025}, though differing nominal grating geometries could also contribute to these variations. Therefore, future work seeking to transition our technology to volume manufacturing while fully benefiting from its narrow linewidth should prioritize developments on device uniformity as opposed to improving inherent material properties. 
    
    Additional fabrication trials performed in this work further substantiate this claim. Namely, we annealed two of the fields shown in Fig.~\ref{fig:fig1}{b} in oxygen at 500~$^\circ$C over two hours in an attempt to remove crystal defects in the lithium tantalate film conducive to photorefractive instability~\cite{katz_Vaportransport_2004a}. However, as statistically shown in Figs.~\ref{fig:fig2}{d-g}, annealed devices did not exhibit noticeably improved linewidths over some of the unannealed ones. 

    In terms of tuning capabilities, the tuning efficiency of the reported lithium tantalate e-DBRs is similar to that of similarly structured lithium niobate ones~\cite{siddharth_Ultrafast_2025}. This similarity arises from their nearly identical $r_{33}$ Pockels coefficients accounting for most of the tuning. Minor differences likely arise from slightly differing nominal waveguide geometries and electrode-to-waveguide spacings, as well as the material permittivity, which affect the reflected mode's electro-optic overlap with the applied electric field and the strength of that field for a given applied voltage, respectively. The frequency response of the Pockels-based tuning, shown in Fig.~\ref{fig:fig3}h, exhibits a 3-dB bandwidth of around \SI{200}{\mega\hertz}. This value is also similar to what was previously reported for similarly structured lithium niobate lasers~\cite{siddharth_Ultrafast_2025}, as expected given their otherwise identical electrode geometry and the similar effective capacitance of the two ferroelectric membranes. Though electro-optic modulation in lithium tantalate PICs can exceed~\SI{100}{\giga\hertz}~\cite{wang_Lithium_2024, lin_Copper_2026, cai_Heterogeneously_2026}, the capacitive nature of the e-DBR's tuning electrodes primarily limits its bandwidth. Altering the geometry of the electrodes to further reduce their capacitance could lead to greater bandwidths. Shorter e-DBRs could also further reduce the capacitance, although this would likely come at the cost of increased linewidth~\cite{tran_Tutorial_2019}.

    Compared to lithium niobate~\cite{holzgrafeRelaxationElectroopticResponse2024}, lithium tantalate PICs can exhibit low modulation drift~\cite{lin_Copper_2026}  attributed to charge carrier drift and photorefractive effects arising from crystalline defect sites. As shown in Figs.~\ref{fig:fig3}e,f and Supplementary Note 1, this work's e-DBR lasers achieve stable emission tuning under Hz-level driving conditions similar to those yielding stable operation in lithium tantalate modulators~\cite{lin_Copper_2026}. This capability further solidifies lithium tantalate's position as a stable material for Pockels-based integrated photonics.
    Nonetheless, tuning drift still eventually arises under very low frequency and amplitude drive voltages, thereby indicating that our devices likely have some carriers that can drift around the waveguide under these tuning conditions. 

    Besides improving process uniformity for increasing lasing performance, future fabrication work should also focus on introducing materials suitable for electronics integration. For instance, relying on lithium tantalate PIC fabrication exploiting the copper Damascene process~\cite{lin_Copper_2026} could introduce prospects for dense co-integration of the laser with microelectronics by means of copper-copper hybrid bonding~\cite{s.k.moore_Copper_2024}. This would considerably increase the scalability of systems-level photonic devices for applications such as LiDAR~\cite{lukashchuk_Photonicelectronic_2024}.

    In summary, we introduced hybrid integrated e-DBR lasers based on lithium tantalate. Wafer-scale measurements reveal that this platform can provide output powers similar to prior integrated e-DBR lasers~\cite{huang_Highpower_2019, xiang_Ultranarrow_2019,siddharth_Piezoelectrically_2024,siddharth_Ultrafast_2025,xue_Pockels_2025} while exhibiting improved linewidths of a few tens of Hz. Tuning ranges and bandwidths line up with prior reports on integrated Pockels lasers~\cite{siddharth_Ultrafast_2025} and straightforward alterations in the geometry of the PIC electrodes can easily improve our reported figures. Provided future improvements on wafer-level device uniformity, the presented metrics along with lithium tantalate's suitability for volume manufacturing~\cite{wang_Lithium_2024} make a compelling case for lithium tantalate e-DBRs as a scalable platform for ultrafast, widely tunable, and narrow linewidth integrated lasers, suitable for emerging applications like LiDAR~\cite{wangHighPerformanceIntegratedLaser2024}, distributed fiber-optic sensing~\cite{luDistributedOpticalFiber2019}, and gas sensing~\cite{cassidyAtmosphericPressureMonitoring1982}.
    
	\begin{footnotesize}
		
		\noindent \textbf{Data and Code Availability Statement}: \blue{The data and code generated in this study have been deposited in the Zenodo database under accession code [https://doi.org/10.5281/zenodo.22652281]~\cite{larocque_zenodo}}.

	\end{footnotesize}
       
    \bibliography{citations}

    \begin{footnotesize}


		\noindent \textbf{Acknowledgments}:
		The PICs were fabricated in the EPFL Center of MicroNanoTechnology (CMi) and IPHYS cleanroom. The LTOI wafers were fabricated in Shanghai Novel Si Integration Technology (NSIT) and the SIMIT-CAS. This work was supported by the EU Horizon Europe research and innovation programme under grant no. 101131069 (Agilight), as well as funding from the Swiss State Secretariat for Education (SERI) and from the Swiss National Science Foundation under grant no. 211728 (Bridge Discovery).

        \noindent \textbf{Competing Interests}:
		T.J.K is a co-founder and shareholder of Deeplight SA, St. Sulpice, Switzerland, a start-up commercializing PIC-based frequency-agile, low noise lasers. The other authors declare no competing interests.

	\end{footnotesize}
    
\end{document}


\title{Supplementary Information for:
\\
Lithium tantalate integrated extended grating-based lasers with Hertz linewidth}

    \author{Hugo~Larocque} \thanks{These authors contributed equally.}
    \affiliation{Institute of Physics, Swiss Federal Institute of Technology Lausanne (EPFL), CH-1015 Lausanne, Switzerland}
    
    \author{Zhuoya~Yuan} \thanks{These authors contributed equally.}
    \affiliation{Institute of Physics, Swiss Federal Institute of Technology Lausanne (EPFL), CH-1015 Lausanne, Switzerland}
    
    \author{Zihan~Li}
    \affiliation{Institute of Physics, Swiss Federal Institute of Technology Lausanne (EPFL), CH-1015 Lausanne, Switzerland}

    \author{Giovanni~Scarioni}
    \affiliation{Institute of Physics, Swiss Federal Institute of Technology Lausanne (EPFL), CH-1015 Lausanne, Switzerland}

    \author{Annika~Schoels}
    \affiliation{Institute of Physics, Swiss Federal Institute of Technology Lausanne (EPFL), CH-1015 Lausanne, Switzerland}

    \author{Jiale~Sun}
    \affiliation{Institute of Physics, Swiss Federal Institute of Technology Lausanne (EPFL), CH-1015 Lausanne, Switzerland}

    \author{Anat~Siddharth}
    \affiliation{Deeplight SA, St-Sulpice CH-1025, Switzerland}

    \author{Xin~Ou}
    \affiliation{State Key Laboratory of Materials for Integrated Circuits, Shanghai Institute of Microsystem and Information Technology, Chinese Academy of Sciences, Shanghai, China}

    \author{Simone~Bianconi}
    \email[]{simone.bianconi@epfl.ch}
    \affiliation{Institute of Physics, Swiss Federal Institute of Technology Lausanne (EPFL), CH-1015 Lausanne, Switzerland}
    
    \author{Tobias~J.~Kippenberg}
    \email[]{tobias.kippenberg@epfl.ch}
    \affiliation{Institute of Physics, Swiss Federal Institute of Technology Lausanne (EPFL), CH-1015 Lausanne, Switzerland}
    \affiliation{Institute of Electrical and Micro engineering, Swiss Federal Institute of Technology, Lausanne (EPFL), CH-1015 Lausanne, Switzerland}

\setcounter{equation}{0}
\setcounter{figure}{0}
\setcounter{table}{0}

\setcounter{subsection}{0}
\setcounter{section}{0}
\setcounter{secnumdepth}{3}

\maketitle
{\hypersetup{linkcolor=blue}\tableofcontents}
\newpage

\section{Low voltage emission tuning measurements}
\label{sec:drift}

\begin{figure*}[htbp!]
	\centering
	\includegraphics[width=1\linewidth]{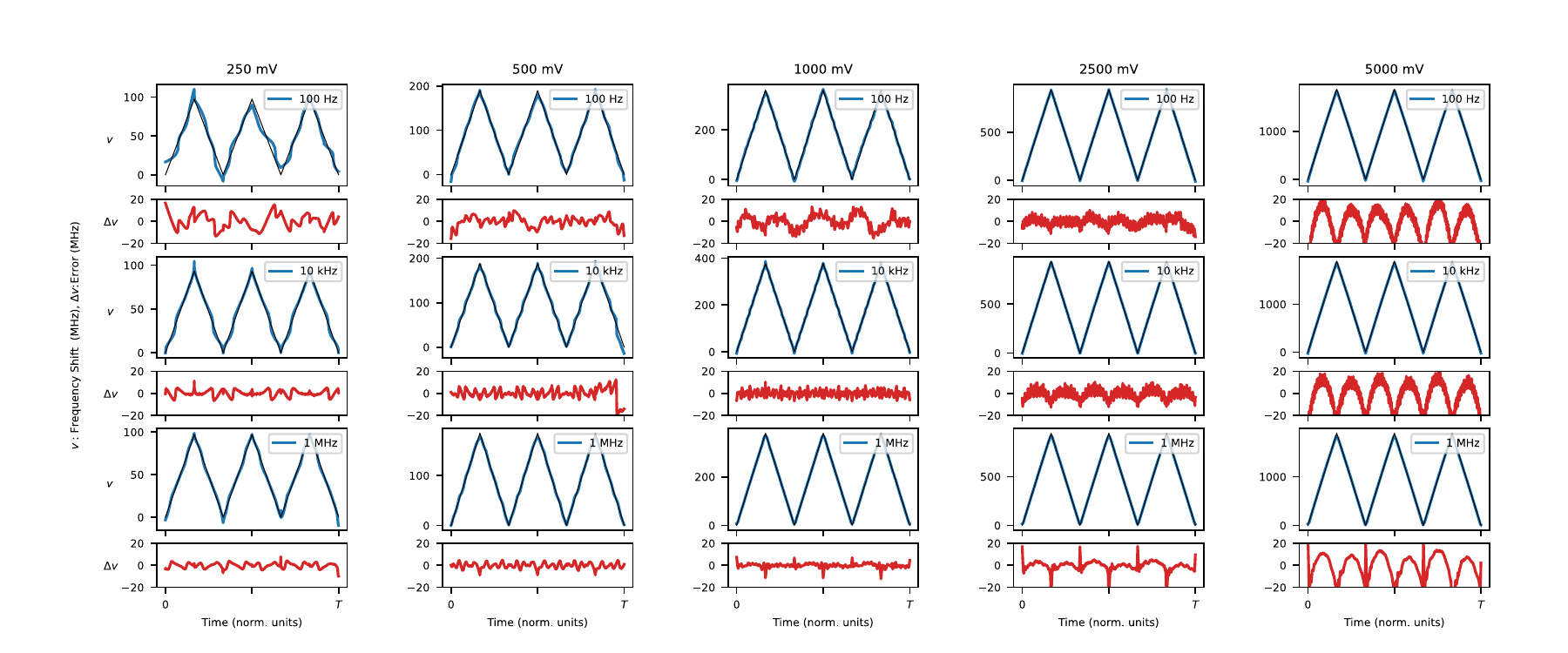}
	\caption{\textbf{Low voltage frequency tuning response of lithium tantalate e-DBR lasers.}
        Measured optical frequency shift in the integrated laser spectrum while applying 0.25~Vpp - 5~Vpp triangle waves to the lithium tantalate grating at \SI{100}{\hertz}, \SI{10}{\kilo\hertz}, and \SI{1}{\mega\hertz} drive frequencies. The laser's grating features a \SI{1.3}{\micro\meter} nominal waveguide width, \SI{375}{\nano\meter} post diameter, \SI{600}{\nano\meter} post-to-waveguide gap, and a \SI{900}{\nano\meter} period. The plotted error signals, $\Delta \nu$, correspond to the difference between the measured data and the perfect triangle waves plotted as black lines.
		}
	\label{figS1}
\end{figure*}

As shown in Figs.~3e,f of the main text, tuning lithium tantalate e-DBR lasers deviates further from its ideal linear behavior as both the frequency and the amplitude of the drive voltage decrease. Supplementary Figure~\ref{figS1} provides additional insight on this trend by plotting the frequency tuning ramps measured while tuning the laser with triangle waves defined by the specified amplitudes and frequencies at room temperature. As drive frequency and amplitude decrease, the linear deviation signal features increasingly irregular features that likely stem from carrier drift known to occur in lithium niobate~\cite{holzgrafeRelaxationElectroopticResponse2024} and tantalate~\cite{lin_Copper_2026}.

\section{Tuning efficiency modeling}
\label{sec:modeling}

As mentioned in the main text, the extended-DBR emission tuning arises from an induced effective index change within the grating's lithium tantalate. Upon applying a voltage on the extended-DBR's electrodes, and electric field changes the material's refractive index due to the Pockels effect. This change in turn manifests in the effective index, $n_\text{eff}$, of the mode propagating through a DBR with a fixed period of $\Lambda$, thereby changing the grating's resonant wavelength as prescribed by the Bragg condition, $\lambda = 2 n_\text{eff} \Lambda$.

\begin{figure*}[htbp!]
	\centering
	\includegraphics[width=1\linewidth]{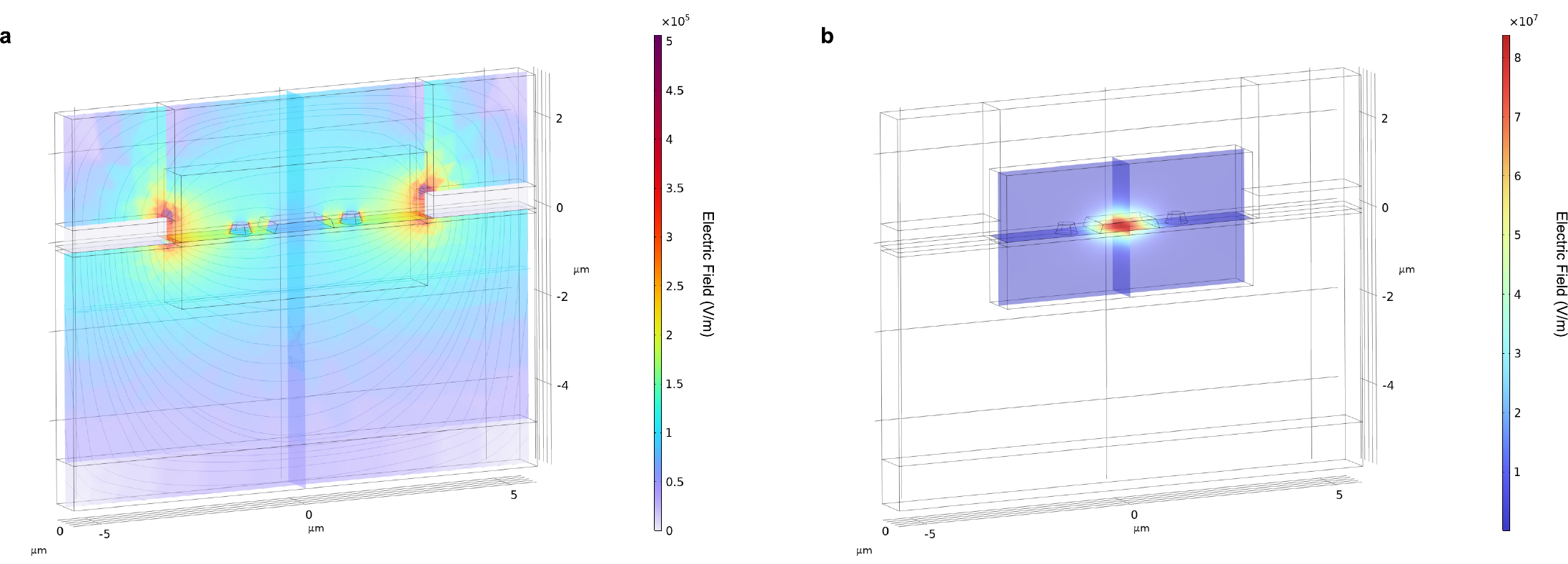}
	\caption{\textbf{Finite element simulations predicting the tuning efficiency of lithium tantalate extended-DBR lasers.} 
        \textbf{a,} Electric field generated by the electrodes onto a unit cell of the DBR upon applying a 1~V DC voltage. \textbf{b,} Optical field of the mode reflected by the lithium tantalate extended-DBR.
		}
	\label{figS2}
\end{figure*}

To simulate this shift, we solve for both the applied electric field under static conditions and the DBR's eigenmode using a Finite element method (FEM) solver (COMSOL multiphysics). Supplementary Figures~\ref{figS2}a,b provide the applied DC electric field and optical mode profile of the perturbed grating, respectively. Based on a perturbative calculation adapted for optical materials experiencing a refractive index shift due to the Pockels effect~\cite{larocque_Photonic_2024, cai_Heterogeneously_2026}, the frequency shift experienced by the eigenmode $E_n$ under the application of an external electric field $E_\text{DC}$ aligned along lithium tantalate's extraordinary axis approximately becomes:
\begin{equation}
    \Delta \omega_n = -\frac{\omega_n}{2}
    \frac{\displaystyle \int_\text{\LT} dV \, E_{\rm{DC}}(\mathbf{r}) \left[r_{13}\epsilon_o^2\left( |E_{n,1}(\mathbf{r})|^2 + |E_{n,2}(\mathbf{r})|^2\right) + r_{33}\epsilon_e^2 |E_{n,3}(\mathbf{r})|^2\right]}{{\displaystyle \int dV \,\epsilon_{11}(\mathbf{r})\, |E_{n,1}(\mathbf{r})|^2 +\epsilon_{22}(\mathbf{r}) |E_{n,2}(\mathbf{r})|^2 +\epsilon_{33} (\mathbf{r}) |E_{n,3}(\mathbf{r})|^2}},
\end{equation}
where we neglect the fringing effects in the applied electric field and take advantage of the diagonal permittivity tensor $\epsilon_{mn}$ of the resulting system.

The interplay between the electro-optically tuned DBR and the gain medium ultimately determines the tuning efficiency of the lithium tantalate e-DBR laser. To simulate such phenomena, we rely on a time-domain model of first-order forward, $E^+$, and reverse, $E^-$, traveling waves propagating between the active gain section and the grating reflector~\cite{gallagherWavelengthTunableLasers2005, t.smyIntegrationTravelingWave2020, j.h.rasmussenImplementationTravelingWave2023}, as previously investigated in works on lithium niobate e-DBR Pockels lasers~\cite{siddharth_Ultrafast_2025}. Herein, propagation in the III-V gain chip verifies:
\begin{equation}
    \frac{1}{v_\text{g,g}}\frac{d E_\text{g}^\pm(z,t)}{dt}-\frac{d E_\text{g}^\pm(z,t)}{dz} = 
    \left(-\frac{\alpha_\text{g}}{2}+G(z,t)\left(\frac{1}{1+\epsilon S}-i\alpha_\text{H}\right)\right)
    \cdot E_\text{g}^\pm(z,t) + F_N^\pm
\end{equation}
where the gain profile $G(z,t)$ and local photon density $S(z,t)$ satisfy
\begin{eqnarray}
    S(z,t) &=& \left| E_\text{g}^-(z,t) \right|^2 + \left| E_\text{g}^+(z,t) \right|^2,\\
    G(z,t) &=& G_0 (N(z,t)-N_0).
\end{eqnarray}
Here, $N(z,t)$ consists of the one-dimensional excited charge carrier number density and follows the optical pumping model~\cite{e_a_avrutin_dynamics_2009}
\begin{equation}
    \frac{d N(z,t)}{dt} = \frac{J}{edh} - N(z,t) (\tau_\text{nr}^{-1} + BN(z,t) + C N^2(z,t)) - v_\text{g,g} S(z,t) G(z,t).
\end{equation}
As for propagation in the lithium tantalate photonic integrated circuit, the following equations respectively account for propagation dynamics within the chip's linear inverse taper and its Bragg grating:
\begin{eqnarray}
    \frac{1}{v_\text{g,t}}\frac{d E_\text{t}^\pm(z,t)}{dt}-\frac{d E_\text{t}^\pm(z,t)}{dz} &=& 
    -\frac{\alpha_\text{g}}{2}
    E_\text{t}^\pm(z,t), \\
    \frac{1}{v_\text{g,b}}\frac{d E_\text{b}^\pm(z,t)}{dt}-\frac{d E_\text{b}^\pm(z,t)}{dz} &=& 
    \left(i\delta -\frac{\alpha_\text{g}}{2}\cdot F_\text{E}\right)
    E_\text{b}^\pm(z,t) + i K(z,t) E_\text{b}^\mp(z,t),
\end{eqnarray}
where the following equations enforce boundary conditions attributed to the III-V lithium tantalate chip interface and reflection from the backside of the gain medium:
\begin{eqnarray}
    E_\text{t}^+(0,t+dt) &=& -t \cdot e^{i\phi} \cdot E_g^+(L_\text{g},t), \\
    E_\text{g}^-(L_\text{g},t+dt) &=& +t \cdot e^{i\phi} \cdot E_t^-(0,t), \\
    E_\text{g}^+(0,t+dt) &=& -r \cdot E_g^-(0,t).
\end{eqnarray}
Here, $t$ corresponds to the coupling transmission between the two chips where we ignore reflections due their suppression from the two chip's angled tapers. $\phi$ consists of the total intracavity phase. 

\begin{table*}[htbp]
\centering
\begin{tabular}{@{\extracolsep{\fill}} P{3cm} P {4cm} P{10cm}  @{}}
\toprule
\textbf{Parameter} & \textbf{Value} & \textbf{Description}\\
\midrule
$I$ &  0-500~mA & Pump current\\
$L_\text{g}$ &  $1\times 10^{-3}$~m & Length of gain section\\
$d$ &  $5\times 10^{-6}$~m & Width of gain section\\
$h$ &  $5\times 10^{-9}$~m & Height of quantum well gain section\\
$v_\text{g,g}$ &  $8.57\times 10^{7}$~m s$^{-1}$ & Gain section group velocity\\
$\alpha_\text{g}$ &  $1\times 10^{3}$~m$^{-1}$ & Gain section passive propagation loss\\
$N_0$ &  $1.2\times 10^{24}$~m$^{-3}$ & Carrier density at transparency\\
$\tau_\text{nr}$ &  $1\times 10^{-9}$~s & Non-radiative carrier lifetime\\
$G_0$ &  $5\times 10^{-4}$~m$^{-1}$ & Differential gain\\
$\epsilon$ &  $1\times 10^{-5}$ & Nonlinear compression factor\\
$B$ &  $1\times 10^{-16}$~m$^3$s$^{-1}$ & Bimolecular recombination coefficient\\
$C$ &  $1\times 10^{-40}$~m$^6$s$^{-1}$ & Auger recombination rate\\
$F_N$ &  $1\times 10^{-22}$~m$^{-2}$ & Spontaneous emission noise term\\
$\alpha_\text{H}$ &  5 & Linewidth enhancement factor\\
e &  $1.6\times 10^{-19}$~C & Electron charge\\
\midrule
$L_\text{t}$ &  $1 \times 10^{-3}$~m & Length of taper coupler section\\
$L_\text{b}$ &  $7.8\times 10^{-3}$~m & Length of grating\\
$v_\text{g,(b,t)}$ &  $1.33 \times 10^{8}$~m\,s$^{-1}$ & Group velocity of grating and taper sections\\
$K$ &  100~m$^{-1}$ & Bragg coupling coefficient\\
$\alpha_\text{b,t}$ &  2~m$^{-1}$ & Grating and taper section passive propagation loss\\
$\delta_\text{b}$ &  0~m$^{-1}$ & Detuning of Bragg grating\\
$t$ &  1 & Facet transmission\\
$\phi$ &  0 & Intracavity phase shift\\
\bottomrule
\end{tabular}
\caption{\textbf{Parameter values used to model electro-optic emission tuning.}}
\label{tab:parameters}
\end{table*}

Supplementary Table~\ref{tab:parameters} provides a list of the parameter values used in our modeling. To solve the above coupled differential equations, we rely on a first order upwind Euler integration scheme with an exponential solution for the phase shift per element~\cite{j.h.rasmussenImplementationTravelingWave2023}. 121, 75, and 544 sections model the gain, taper, and Bragg grating sections, respectively. The integration carries over a $dt=100$~fs time step enforcing numerical stability as prescribed by the condition $dz_\text{g,b}\geq v_\text{g,(g,b)}\, dt$. Supplementary Figure~\ref{figS3} provides the results of these simulations. The results of Supplementary Figure~\ref{figS3}a suggest that devices with weaker grating strengths results in higher output powers with mode hops expected to occur near 100~mA and 300~mA pump currents. These results confirm that weaker reflections result in weaker feedback reflected back into the gain material, which correspondingly allocates more power in the laser's output.

Supplementary Figure~\ref{figS3}b gives the tuning range of the corresponding lasers while detuning the grating's response by 0.78~GHz at a rate of 30~MHz. Based on the finite element simulations outlined in Supplementary Figure~\ref{figS2}, this detuning corresponds to 1~V applied to the lithium tantalate e-DBR. The resulting laser tuning roughly corresponds to 50\% of the grating's tuning. This fraction lines up with the relative group delays in the gain and grating chips forming the hybrid laser cavity~\cite{siddharth_Ultrafast_2025}. Specifically, the global phase shift in the hybrid cavity remains constant while the laser is in operation. Therefore, shifts in the laser's emission frequency will compensate any electro-optically induced shifts in the grating response. The group delay attributed to the lithium tantalate chip roughly corresponds to 50\% of the hybrid cavity's total group delay, thereby causing the resulting detuning of the cavity to be 50\% smaller than that of the grating. The achieved tuning further reduces for e-DBRs with stronger grating strengths. This trend agrees with the above group delay discussions. Stronger reflections correspond to shorter propagation through the e-DBR, thereby reducing its contributions to the hybrid cavity group delay and hence the overall tunability of the laser. However, this trend reverses at very weak grating strengths and higher pump currents, thereby indicating potentially more complex lasing dynamics caused by the interplay between the gain medium and lower feedback powers.

Supplementary Figures~\ref{figS3}c,d provide the corresponding nonlinearity and residual amplitude modulation figures while detuning the lithium tantalate e-DBR.

\begin{figure*}[htbp!]
	\centering
	\includegraphics[width=0.75\linewidth]{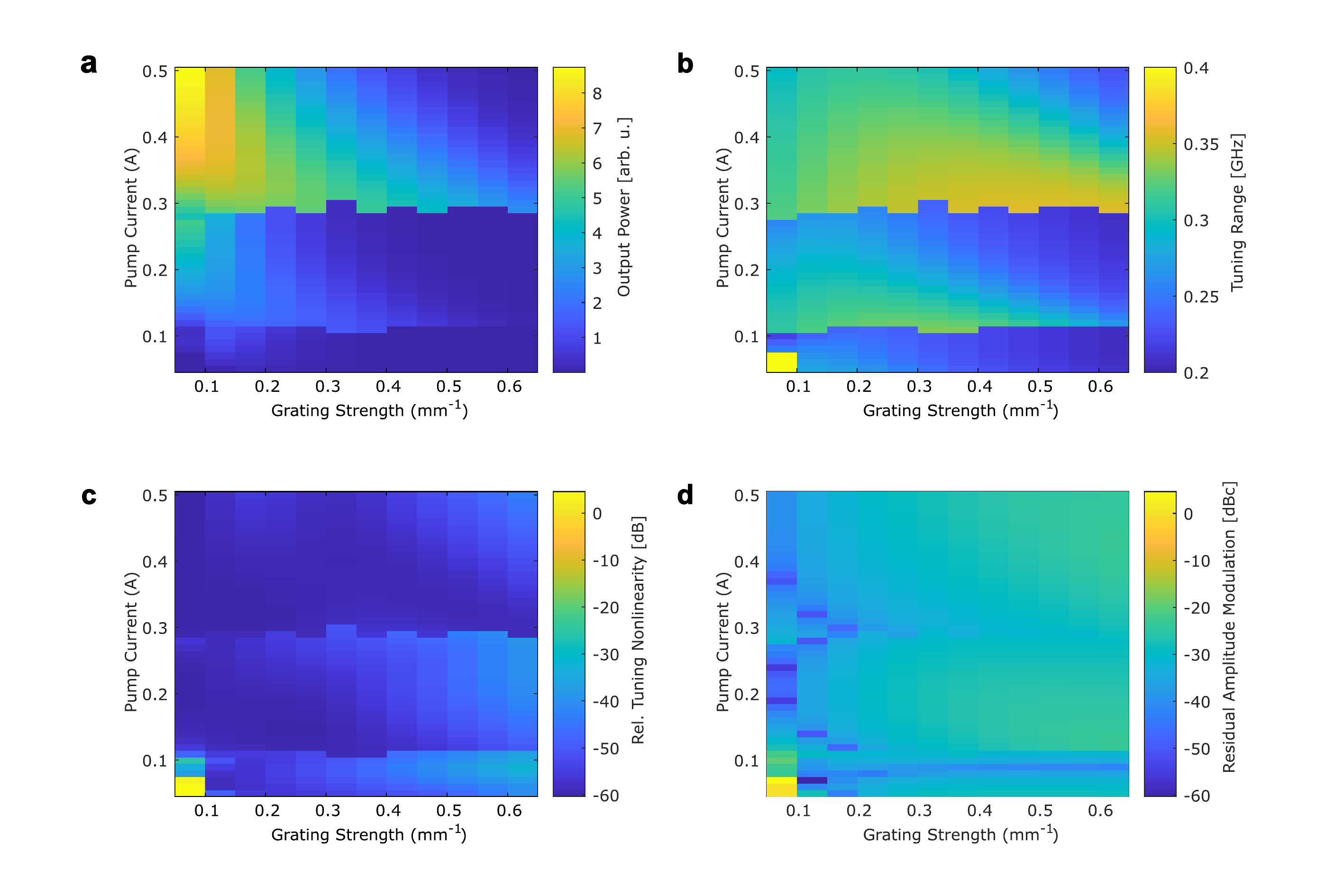}
	\caption{\textbf{Traveling wave simulations of lithium tantalate extended-DBR lasers.} 
        \textbf{a,} Relative output power of the e-DBR laser for various grating strengths and pump currents for a device defined by the parameters in Supplementary Table~\ref{tab:parameters}. \textbf{b,} Corresponding tuning range under a 2~Vpp and 30~GHz drive signal assuming a grating tuning sensitivity of 0.78~GHz/V as determined by the methodology from Supplementary Figure~\ref{figS2}. \textbf{c,} Corresponding relative tuning nonlinearity and \textbf{d,} residual amplitude modulation under the same driving conditions considered in \textbf{b}.
		}
	\label{figS3}
\end{figure*}


\section{Packaging of lithium tantalate extended-DBR lasers}

The lithium tantalate e-DBR laser is packaged to improve its mechanical stability and enable portable operation beyond a bench-top alignment setup. As shown in Supplementary Figure~\ref{figS4}, an active packaging process is used to assemble the photonic integrated circuit (PIC), reflective semiconductor optical amplifier (RSOA), and fiber array unit (FAU). The FAU is first aligned and fixed to provide a stable output interface. The heights of the RSOA submount and the PIC spacer are then carefully matched to enable efficient and stable edge coupling between the RSOA and the PIC. The specific geometry parameters used in the laser packaging are shown in Supplementary Table~\ref{tab:geometry}.

\begin{figure*}[htbp!]
	\centering
	\includegraphics[width=0.75\linewidth]{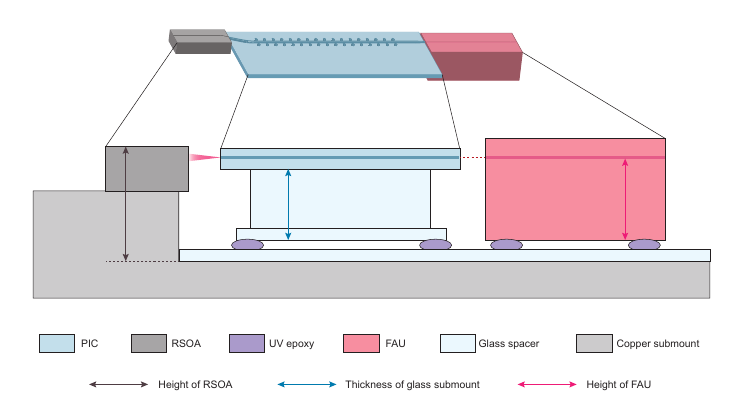}
	\caption{\textbf{Schematic of packaging geometry.}}
	\label{figS4}
\end{figure*}

\begin{table*}[htbp]
    \centering
    \begin{tabular}{@{\extracolsep{\fill}} P{4.5cm} P {2.5cm} P{10cm}  @{}}
    \toprule
    \textbf{Parameter} & \textbf{Value} & \textbf{Description}\\
    \midrule
    PIC thickness & 240 ~\textmu m & Thickness of the PIC, including top cladding\\
    Thickness of glass submount & $\sim$ 1260 ~\textmu m & Total thickness of the glass submount under the PIC\\
    Height of FAU &  1500 ~\textmu m & The height from FAU waveguide to the bottom of FAU\\
    Height of RSOA & $\sim$ 1700 ~\textmu m & Total height including the ridge of the copper submount and RSOA\\
    Epoxy gap & 40-100 ~\textmu m & Estimated thickness of UV epoxy\\
    \bottomrule
    \end{tabular}
    \caption{\textbf{Geometry parameters of the main components used in integrated laser packaging.}}
    \label{tab:geometry}
\end{table*}
    
The RSOA and PIC are actively aligned by monitoring the output power. After optimizing the coupling, UV-curable epoxy is applied to the bottom and facet-side regions for mechanical fixation. The position, volume, and curing sequence of the epoxy are carefully controlled to reduce cure-induced displacement and minimize coupling drift during packaging.

\section{Relative intensity noise (RIN) measurement of packaged extended-DBR laser}

Relative intensity noise (RIN) is defined as the power-noise spectral density normalized by the square of the mean optical power, thereby featuring the relative fluctuation of optical power. We measure the RIN of the extended-DBR laser with an OEWaves OE4000 laser phase noise measurement system.



\begin{figure*}[htbp!]
	\centering
	\includegraphics[width=0.65\linewidth]{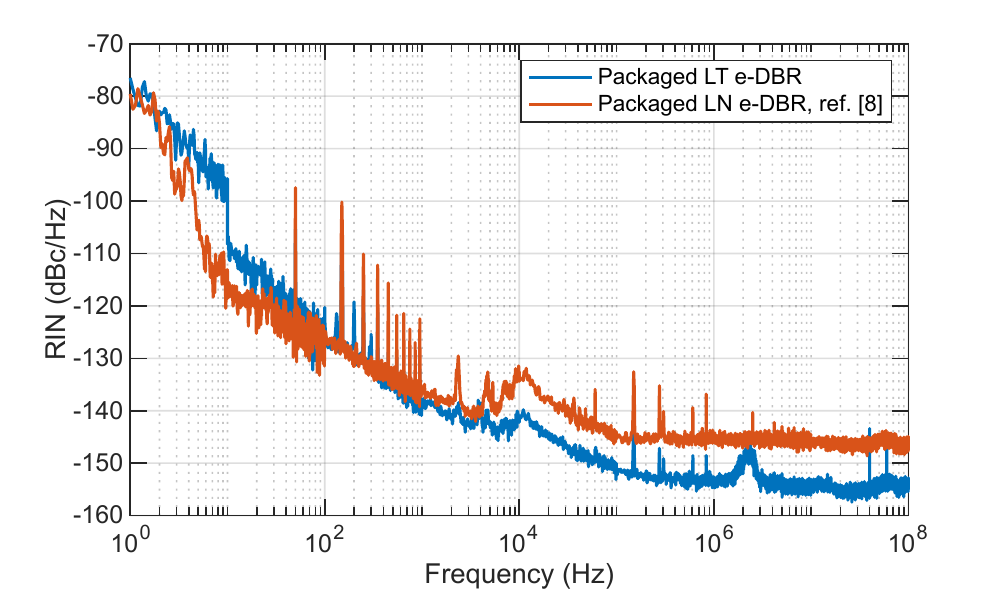}
	\caption{\textbf{Relative intensity noise characterization of extended-DBR lasers.} Abbreviations: LT: lithium tantalate, LN: lithium niobate}
	\label{figS5}
\end{figure*}

The relative intensity noise of the packaged lithium tantalate e-DBR integrated laser is shown in Supplementary Figure~\ref{figS5}. For reference, we also provide results for a packaged lithium niobate e-DBR laser developed in prior work~\cite{siddharth_Ultrafast_2025}. Both the packaged lithium tantalate e-DBR laser and lithium niobate e-DBR laser have similar noise level at low-frequency offsets, which is dominated by thermal noise, TCCR and technical noise~\cite{zhang2025fundamental}. At high frequency offsets, the lithium tantalate e-DBR laser exhibits a reduction in RIN compared with the lithium niobate e-DBR laser, approaching the shot noise or electronic noise floor.

\section{Light detection and ranging}

\begin{figure*}[htbp!]
	\centering
	\includegraphics[width=0.99\linewidth]{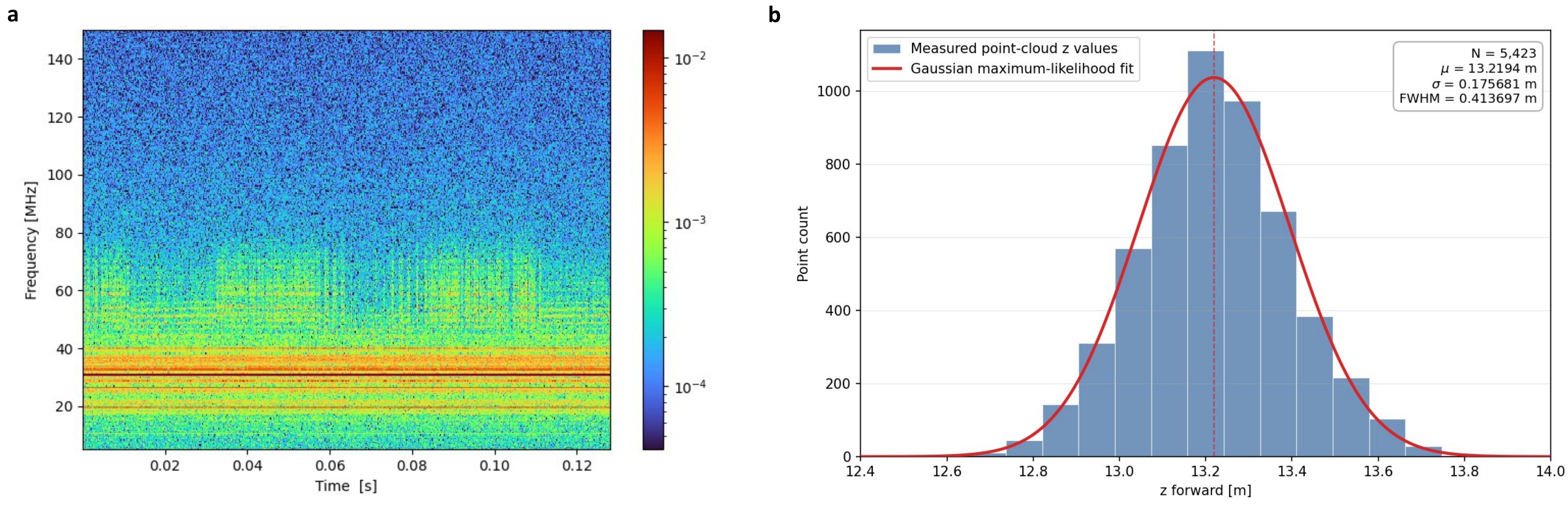}
	\caption{\textbf{Acquired LiDAR data using the lithium tantalate extended-DBR laser.} \textbf{a}, Downsampled interferogram obtained from the LiDAR data. \textbf{b}, Histogram showing the distribution of the calculated range of the background flat surface in the point cloud shown in Fig.~4g of the main text.}
	\label{figS6}
\end{figure*}

We exploited the fast tunability of the lithium tantalate e-DBR laser to conduct a proof-of-concept optical coherent ranging experiment using the FMCW LiDAR scheme using the setup shown in Fig.~4f of the main text. The laser was chirped using a triangular ramp with a 50 kHz repetition rate, obtaining a 1.65 GHz tuning range. Supplementary Figure~\ref{figS6}a shows the interferogram obtained from the measurement. The beat note signals between 15 and 40 MHz are due to the reflections of the collimator and of the circulator in the setup. The signals between 45 and 75 MHz are due to the targets shown in the inset of Fig.~4g of the main text. In each of the time slices of the interferogram, the peak corresponding to the target reflection is selected and converted to range. The range resolution is extrapolated by fitting a Gaussian to the histogram of the calculated range of a flat surface. Supplementary Figure~\ref{figS6}b shows such a histogram for the flat background surface in the point cloud of Fig.~4g of the main text. The range resolution of our measurement obtained from the Gaussian fit is below 18 cm. 

\section{Comparison with other extended-DBR lasers}

Supplementary~Table~\ref{tab:comparison} compares various performance metrics of this work's e-DBR lithium tantalate laser against e-DBR lasers implemented in other integrated photonic material platforms.

\begin{table*}[htbp]
    \centering
    \setlength{\tabcolsep}{4.2pt}
    \renewcommand{\arraystretch}{1.12} 
    \footnotesize
    \begin{tabular*}{\textwidth}{@{\extracolsep{\fill}} c c c c c c @{}}
    \toprule
    \textbf{Ref.} & 
    \textbf{Platform} &
    \textbf{Intrinsic linewidth} &
    \textbf{$\beta$ linewidth} &
    \textbf{Output power} &
    \textbf{Integration Approach}
    \\
    \midrule
    This work & {\LT} & \SI{530}{\hertz} (rep.) & \SI{12}{\kilo\hertz} (rep.)  & \SI{4}{\milli\watt} (rep.) & Hybrid\\
     & &  \SI{24}{\hertz} (min.) & \SI{1.9}{\kilo\hertz} (min.) & \SI{10}{\milli\watt} (max.) & Hybrid\\
     \cite{siddharth_Ultrafast_2025} & {\LN} & \SI{2.8}{\kilo\hertz}& \SI{191.5}{\kilo\hertz} & \SI{12.5}{\milli\watt} & Hybrid\\
     \cite{xue_Pockels_2025} & {\LN} & \SI{167}{\hertz}& not reported & \SI{13}{\milli\watt} & Hybrid \\
     \cite{li_integrated_2022} & {\LN} & \SI{11.3}{\kilo\hertz}& not reported & \SI{5.5}{\milli\watt} (on-chip, max.) & Hybrid\\
     \cite{siddharth_Piezoelectrically_2024} & {\SiN} &  \SI{2.5}{\kilo\hertz}& \SI{24.2}{\kilo\hertz} & \SI{25}{\milli\watt} & Hybrid\\
     \cite{xiang_Ultranarrow_2019} & {\SiN} & \SI{320}{\hertz}& not reported & \SI{24}{\milli\watt} & Hybrid\\
     \cite{huang_Highpower_2019} & {\Si} & \SI{1.1}{\kilo\hertz}& not reported & \SI{37}{\milli\watt} (on-chip) & Heterogeneous\\
    \bottomrule
    \end{tabular*}
    \caption{\textbf{Figures of merit achieved in integrated e-DBR lasers.} Abbreviations: rep.: representative, min.: minimum, max.: maximum.}
    \label{tab:comparison}
    \end{table*}

\bibliography{citations}